\documentstyle[preprint,aps]{revtex}  
\draft
\title{Elastic critical behaviour in a $3d$ model for polymer gels}
\author{Emanuela Del Gado,$^{a,c}$ Lucilla de Arcangelis, $^{b,c}$ and
        Antonio Coniglio $^{a,c}$}
\address{${}^a$Dipartimento di Scienze Fisiche, Universit\`a di Napoli
         "Federico II",\\ Complesso Universitario di Monte Sant'Angelo,
         via Cintia 80126 Napoli, Italy}
\address{${}^b$Dipartimento di Ingegneria dell'Informazione, 
         Seconda Universit\`a di Napoli, via Roma 29,
 81031 Aversa (Caserta), 
         Italy}
\address{${}^c$ INFM Udr di Napoli and Gruppo coordinato SUN}
\date{today}
\begin{document}
\maketitle
\begin{abstract}
The elastic response in polymeric gels is studied by means of a percolation
dynamic model. By numerical simulations the fluctuations in the 
gyration radius and in the center of mass
motion of the percolating cluster are determined. Their
scaling behaviour at the gelation threshold
gives a critical exponent for the elastic modulus 
$f\sim 2.5 \pm 0.1$ in agreement with the prediction $f=d\nu$.\\
PACS: 05.20.-y, 82.70.Gg, 83.10.Nn
\end{abstract}
%
%
%%%%%%%%%%%%%%%%%%%%%%%%%%%%%%%%%%%%%%%%%%%%%%%%%%%%%%%%%%%%%%%%%%%%%
%
\newpage
\section{Introduction}
The elastic response in polymer gels is due to the presence of the polymeric
network formed in the gelation process: this random structure makes the system 
 able to respond to external stresses. Due to the large 
possibility of conformational changes which characterizes this macromolecular 
phase, there is an entropic term contributing to the elastic properties of the 
system. 
This picture is typical of gelling systems and may fit a wide range of very 
different materials. Then because of the different role played by the entropic 
term and  the particular features of the polymeric networks, a rich
phenomenology is produced.\\ 
The viscoelastic behaviour of gelling materials and the mechanisms at its
origin certainly represent a central issue in soft matter physics and 
have a high relevance to a wide range of applications from food processing
to materials science. On the other hand these materials show many typical 
features of complex systems and their study is
connected to some fundamental problems in soft matter physics,
from the entropic elastic behaviour, which we are interested
in here, to glassy dynamics. 
As a consequence these systems are intensively investigated both 
experimentally and
by means of statistical mechanics models, bringing up a lively debate.\\ 
In particular here we study the critical behaviour of the elastic response 
 of the system as the gelation process takes
place producing the polymeric network: 
in a gelling system the elastic modulus starts growing at the gelation
threshold with a power law
behaviour, usually expressed in terms of the
polymerization degree. In the experiments performed on different gelling
materials the value of the critical exponent $f$ describing the critical 
behaviour of the elastic modulus appears to be close to either $f\sim2$, or  
$f\sim 3$, or else $f\sim4$. The value $f\sim2$ is actually observed in
experiments on agarose gel \cite{aga}, gelatin gels \cite{gel}, some silica 
gels (TEOS) \cite{teos}. Within statistical mechanics models it corresponds to 
the prediction based on 
the de Gennes' analogy between the elasticity of a percolating random 
network of Hookean 
springs and the conductivity in a random percolating network of resistors 
\cite{deg76}. 
The critical exponent $f\sim 3.$ has been observed in diisocianate/triol gel 
\cite{diis}, in epoxy resins \cite{epo}, TMOS silica gels \cite{tmos}, polyesters gels \cite{poly}. It is very close to the value $f=d\nu$, where $d$ is the 
dimensionality of the system and $\nu$ is the critical exponent of the 
connectedness length $\xi$,  predicted 
in ref.\cite{conda}
which in $3d$ gives $f\sim 2.64$ (in mean field both $d\nu$ and the random
resistor network exponent are equal to $3$). Following the argument 
in ref.\cite{conda} this exponent would then 
correspond to the case of a dominating entropic term, evaluated by means of
scaling arguments for the percolating network: the elastic energy of
the system is due to the entropic elasticity of the macrolink 
whose length is of the order of the connectedness correlation length.
Finally the value $f \sim 4$ can be linked to the prediction of the
bond-bending model $f= d\nu +1$ \cite{fense,kawe} or alternatively $f=t+2\nu$ 
\cite{fense2,staubb}, 
which in $3d$ give $f \sim 3.7$. 
This would imply that within the elastic energy describing the system 
there is a bending term playing a relevant role in the elastic response of the
network. This critical behaviour is typical of some colloidal gels \cite{coll}. 
The clustering of the experimental values for the elasticity critical exponent 
in gels around discrete values suggests the possibility of correspondently 
individuating different universality classes,
which should be characterized by some intrinsic features of the networks 
formed in the materials.\\
Recent numerical studies via molecular dynamics \cite{pli1} and 
Monte Carlo simulations \cite{faka} of 
percolating networks of tethered 
particles with no hard core interactions have shown that  
the shear modulus critical exponent is $\sim 1.3$ in $d=2$ and $\sim 2.$ 
in $d=3$. Monte Carlo simulations of   
two and three-dimensional percolating networks of tethered
particles with hard core repulsion \cite{faka} find consistent results for 
the shear modulus critical
exponent. These results agree with the de Gennes prediction 
and with some experimental results.\\ 
Within the numerical studies we have approached the study of 
this problem introducing a percolation dynamic
model, and directly investigating the dynamic viscoelastic properties as the 
percolation transition takes place. 
Our model introduces in the percolation
model the bond-fluctuation dynamics, which takes into account 
the conformational changes of the polymer molecules and the excluded volume
interactions. This model has been translated in a lattice algorithm and studied
via numerical simulations on hypercubic lattices. Actually it presents many
fundamental features of the gelation phenomenology and has already allowed to 
study the
critical behaviour of the viscoelastic properties and the relaxation process in
gelling systems in the sol phase \cite{dedecon3}. Numerical
simulations of the model in $d=2$ \cite{dedecon2} have shown that the
elasticity critical exponent is $\sim 2.7$, a value which agrees with the
prediction $f=d\nu$ in \cite{conda}. The study of the
model in $d=3$ is of great interest to check the elasticity critical
behaviour and to eventually compare the findings with the experimental results.
We here present the
results of large scale $3d$ simulations: the data show that 
$f \sim 2.5 \pm 0.1$, again in agreement with the prediction $f=d\nu$, 
coherently with the findings in $d=2$.\\ 
In the next section we present the model and the details of the numerical
simulations, in section III the elastic response of the percolating cluster is 
discussed and a scaling behaviour is obtained; in section IV the results of the
numerical simulations are
presented and discussed; section V
contains concluding remarks.
\section{Model and numerical simulations}
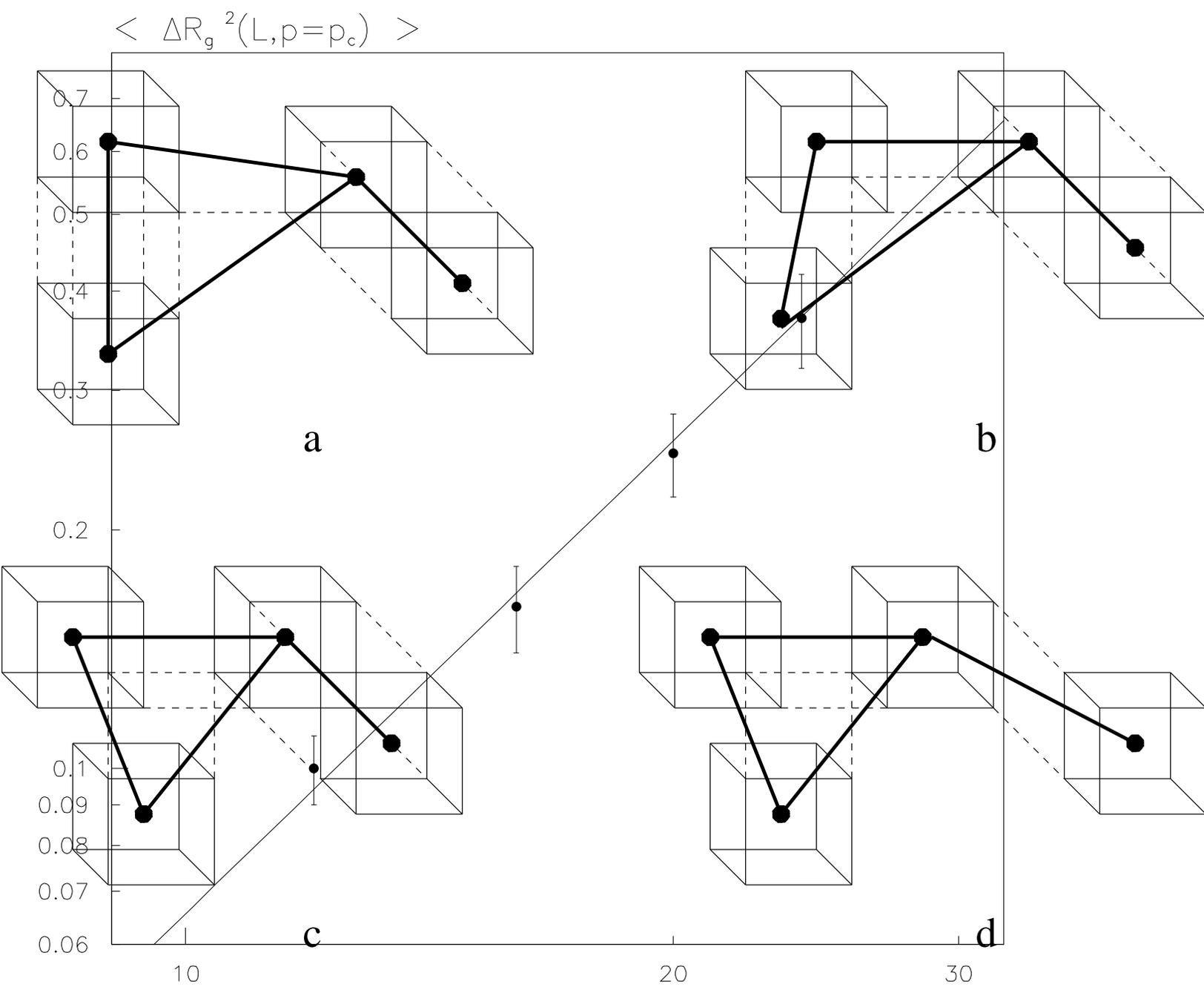
\begin{figure}
\caption{An example of time evolution of a cluster formed by four monomers
according to the bond-fluctuation dynamics: in $a$, starting from the upper
central bond and clockwise, the bond lengths are $l=\sqrt{5},3,3,2$; in $b$ the
upper left monomer has moved forward and $l=2,3,3,\sqrt{5}$; having moved right
 the other left monomer in $b$ one has $c$ with $l=2,3,\sqrt{6},\sqrt{6}$;
moving right the front monomer in $c$ the $d$ configuration is obtained with
$l=2,\sqrt{10},\sqrt{6},\sqrt{6}$.}
\label{fig1}
\end{figure}

We consider a solution of tetrafunctional monomers of concentration $p$. 
The monomers interact via excluded volume interactions, i.e. a monomer occupies
a lattice elementary cell and two occupied cells cannot have common sites. 
Nearest neighbours or next-nearest neighbours are instantaneously linked 
by a permanent bond with probability $p_{b}$. In terms of these two parameters 
the percolation line can be determined via the critical behaviour of the
percolation properties, individuating the sol and the 
gel phase in the system. Actually in the simulation we fix $p_{b}=1$
and study the system varying $p$ \cite{dedecon3,dedecon2,dedecon1}. The
percolation quantities critical exponents agree with the random percolation
predictions \cite{staul} (e.g. $\gamma \simeq 1.8 \pm 0.05$ and $\nu \simeq
0.89 \pm 0.01$ in $3d$ \cite{dedecon3}).   
The monomers free or linked in clusters diffuse via
random and local movements on the lattice according to the bond-fluctuation
dynamics \cite{carkr}, which is ruled by the possibility of varying the bond
lengths within a set of values determined by the excluded volume interactions
and the SAW condition. This produces a high number of different bond vectors
and we consider the case of permanent bonds, which corresponds to the strong
gelation process. In $3d$ the allowed bond lengths on the cubic lattice are 
$l=2,\sqrt{5},\sqrt{6},3,\sqrt{10}$ and in Fig.\ref{fig1} an example of 
different allowed configurations for a polymer molecule is shown.\\ 
We here present the results of extended numerical simulations of the model in 
the gel phase to study the elastic response in the system. 
It is worth noticing that there is no elastic potential energy for the bond 
vectors and then the elastic behaviour is purely entropic implying that our
study is performed at finite temperature.\\ 
The data presented here refer to lattice sizes $L$ ranging from $12$ to $32$, 
and have been averaged over $30$ different realizations.  
The simulations have been performed on the $CRAY-T3E$ system at CINECA taking
 more than $30000$ hours/node of CPU time.
\section{Elastic response in the gel phase}
We study the elastic response in the gel phase in terms of the macroscopic
elastic constant of the system $K$, which is experimentally
defined as the ratio between an applied external force and the deformation.   
In a simple elongation experiment if $l_{0}$ is the undeformed length and 
$\delta = (l- l_{0})$ is the deformation in the 
system, within the linear response approximation the elastic free energy 
$F \sim K \delta^{2}$.
In terms of the Young elastic modulus $E$ the free energy per unit volume is 
$F/V \sim E \delta^{2}/\l_{0}^{2}$. Then $K \sim E V / l_{0}^{2}$ and 
for a cube of size $L$, $K \sim E L^{d-2}$, expressing the fact that the
elastic modulus has the dimensions of an energy per unit volume and is an
intensive quantity whereas $K$ depends on the system size $L$.\\ 
In the gel, since $E$ vanishes at $p_{c}$ as $\sim \xi^{-\tilde{f}}$ (where
$\tilde{f} =f/\nu$) one has
\begin{equation}
 K \sim L^{d-2} \xi^{-\tilde{f}}
\label{eq0}
\end{equation} 
Following eq.(\ref{eq0}) the macroscopic elastic constant presents the 
corresponding scaling behaviours 
as function of the system size $L$ and of the distance from the percolation 
threshold $(p-p_{c})$:
\begin{equation}
\begin{array}{lllll}
p > p_{c}  & & & & 
K \sim L^{d-2}\\
\label{eq1}
p = p_{c} & & & & 
K \sim L^{-\tilde{z}}\\
\label{eq2}
\textrm{fixed}  \hspace{0.4cm} L  & & & &
K \sim (p-p_{c})^{f}
\label{eq3} 
\end{array}
\end{equation}
where $\tilde{z}=\tilde{f}-(d-2)$.\\
An alternative way to obtain these scaling relations is to consider the 
percolating cluster as a network of nodes connected by macrolinks of linear 
size $\xi$ \cite{note}. Each macrolink can be considered as a spring with an 
effective elastic constant. At the percolation threshold there is only one 
macrolink spanning the system and its effective elastic constant coincides 
with the system macroscopic elastic constant $K$.\\ 
In order to evaluate $K$ we notice that  
for a spring of elastic constant $K$ in a thermal bath at temperature $T$, the 
mean fluctuation in the energy $U$ is $\langle \Delta U \rangle = \frac{1}{2} 
K \langle x^{2} \rangle$, where $x$ is the spring elongation. From the energy 
equipartition $K \langle x^{2} \rangle = k_{B} T$, so that at 
the equilibrium the elastic constant $K$ is related to the     
the fluctuations in the spring length \cite{patl,ferryl,doed,trelo}. 
This result can be more generally obtained by means of the Fokker-Plank 
equation for the probability distribution of the spring elongation 
\cite{patl}.\\ 
Therefore the macroscopic elastic constant of the gel phase is 
related to the fluctuations of 
the linear size of the infinite cluster, i.e. 
the squared fluctuations in the gyration radius of the percolating cluster 
$\langle \Delta R_{g}^{2} \rangle = \langle (R_{g} -\langle R_{g} \rangle)^{2} 
\rangle$, $K \sim 1/\langle \Delta R_{g}^{2} \rangle$.\\ 
An alternative way to calculate the macroscopic elastic constant in the gel 
phase by means of the fluctuations in the unperturbed system is to consider the 
center of mass of the percolating cluster as a brownian
particle, subject to a restoring force responsible for the elastic
behaviour of the system. The restoring force 
introduces a limitation on the diffusive process of
a brownian particle. Using the same argument based on the energy equipartition, 
the asymptotic equilibrium value $\Delta$ of its displacement fluctuations 
$\langle \Delta R^{2} (t) \rangle$ 
is inversely proportional to the elastic constant, i.e. $\Delta \sim 1/K$.
\section{Results}
In order to numerically study the elastic response we have used both the 
approaches mentioned before namely we have calculated the 
average fluctuation of the gyration radius of the percolating cluster 
$\langle \Delta R_{g}^{2}\rangle$
and the asymptotic
value $\Delta$ of the mean square
displacement of its center of mass $\langle
\Delta R^{2}(t) \rangle$ \cite{note2}.\\
In the first approach the average fluctuation in the gyration radius
$\langle \Delta R_{g}^{2} \rangle$ of the percolating cluster has been
computed at the percolation threshold in system of different size $L$ with
periodic boundary conditions. 
In Fig.\ref{fig2}
$\langle \Delta R_{g}^{2} \rangle$ at $p_{c}$ is shown in a log-log plot as
function of the lattice size $L$. In the
considered range the data are well fitted by a power law behaviour according 
to eq.(\ref{eq2}), giving a critical exponent
$\tilde{z} \sim 1.9 \pm 0.1$, i.e. $z \sim 1.7 \pm 0.1$. 
As $\tilde{z}=\tilde{f}-(d-2)$, 
$f = z + (d-2)\nu$ and this result gives $f \sim 2.6 \pm 0.1$.\\
In the second approach we have calculated
$\langle \Delta R^{2}(t) \rangle$ at different steps of the
gelation process, i.e. as $p$ grows above the percolation threshold
$p_{c} \sim 0.718 \pm 0.005$ \cite{dedecon3}, and
at $p_{c}$ for different lattice sizes. In these simulations hard-wall 
boundary conditions have been used \cite{note3}. In the gel at the
gelation transition the percolating
cluster is a quite loose network, the center of mass is rather free and the
elastic response is weak. As the gelation process goes on, the network tightens
becoming more rigid, the elastic constant of the system increases and the
center of mass motion is progressively constrained.
This would be then the physical mechanism producing the critical behaviour of
the elastic response for a critical gel. In agreement with our
picture $\langle \Delta R^{2}(t) \rangle$ grows with time up to a limiting
plateau value $\Delta$ as it is shown in Fig.\ref{fig3}.  
This quantity is
inversely proportional to the elastic constant of the
system $\Delta \sim 1/K$ and 
increases as the percolation threshold is approached from above.\\ 
In Figures \ref{fig4},\ref{fig5} and \ref{fig6} the scaling behaviour 
obtained for
$\Delta$ is presented.\\ 
In Fig.\ref{fig4} $\Delta(L,p)$ for $p>> p_{c}$
($p=0.85$) is shown in a
double logarithimic plot: being far from the percolation threshold the system
is reasonably homogeneous and in fact the
scaling behaviour as $\Delta \sim
L^{-0.99 \pm 0.1}$ is observed, in agreement with the behaviour 
$K \sim L^{d-2}$.\\
In Fig.\ref{fig5} $\Delta(L,p=p_{c})$ is shown
in a double logaritmic plot as a function of the lattice size $L$:
the data exhibit a behaviour
$\Delta \sim
L^{\tilde{z}}$ with $\tilde{z} \sim 2.0 \pm 0.1$. This result
again gives $f \sim 2.6 \pm 0.1$.\\ 
Finally in Fig.\ref{fig6}
$\Delta(p-p_{c})$ is plotted for the lattice size $L=32$:
we fit the data with a power law behaviour (eq.(\ref{eq3})) and obtain a
critical exponent $f \sim 2.5\pm 0.1$.\\
It is straightforward to notice that all these numerical
results can be coherently interpreted in terms 
 of the scaling relations obtained for $K$.\\
The value of the
critical exponent $f\sim 2.6$ is in good agreement with the prediction $f=d\nu$
of ref.\cite{conda}, therefore supporting the picture there proposed, and
consistent with the value obtained in the $2d$ study of the model
\cite{dedecon2}.\\
Due to the limited extension of the critical parameter $(p-p_{c})$ and of $L$ 
here investigated the eventual occurrence of a crossover to a different exponent
cannot be excluded.
\section{Conclusions}
The numerical results of Figures \ref{fig2},\ref{fig5} and \ref{fig6} 
show that $\tilde{z} = 2.$, coherently agreeing with 
the prediction $f=d\nu$. On the whole they support the scaling picture we 
propose and the argument of ref.\cite{conda}. They also agree with some
experimental results \cite{diis}-\cite{poly}.  
This result has been obtained via two independent calculations giving
consistent numerical values and is also consistent with the value previously
obtained in the $2d$ study.\\ 
On the other hand the recent numerical works on entropic elastic models of
refs.\cite{pli1,faka} find a good agreement with the de Gennes prediction. 
These results, together with the experimental data, seem to suggest the 
possibility that there are two distinct
universality classes characterized one by an exponent $f=d\nu$ and another by
the electrical analogy exponent $f=t$, which in $3d$ are respectively $\sim
2.64$ and $\sim 2.$. However since the models in the different numerical
studies are rather similar the possibility of a crossover between different 
dynamic regimes, as it is observed in some experiments \cite{teos}, 
cannot be completely excluded.
In both cases these results give a hint for the interpretation of the 
experimental data and indicate the aspects which should be further 
investigated.\\
This work has been partially supported by the European TMR Network-Fractals
under contract No.FMRXCT980183, by INFM PRA-HOP 99, by MURST-PRIN-2000, 
by the INFM Parallel Computing Initiative and by the European Social Fund.      
%
%%%%%%%%%%%%%%%%   REFERENCES  %%%%%%%%%%%%%%%%%%%%%%%%%%%%%%%%%%%%%%%%
%

%
%%%%%%%%%%%%%%%%%%%%%%%%%%%  FIGURE CAPTIONS  %%%%%%%%%%%%%%%%%%%%%%%%%%%%%%

\begin{figure}
\caption{Log-log plot of the fluctuation of the percolating cluster
gyration radius $\langle \Delta R_{g}^{2} \rangle$ as a function of the lattice
size $L$ at $p_{c}$: from the fit of
the data the critical exponent $\tilde{z} \sim 1.9 \pm 0.1$ is obtained. Here
and in the following figures the
lengths are expressed in units of lattice spacing.}
\label{fig2}
\end{figure}

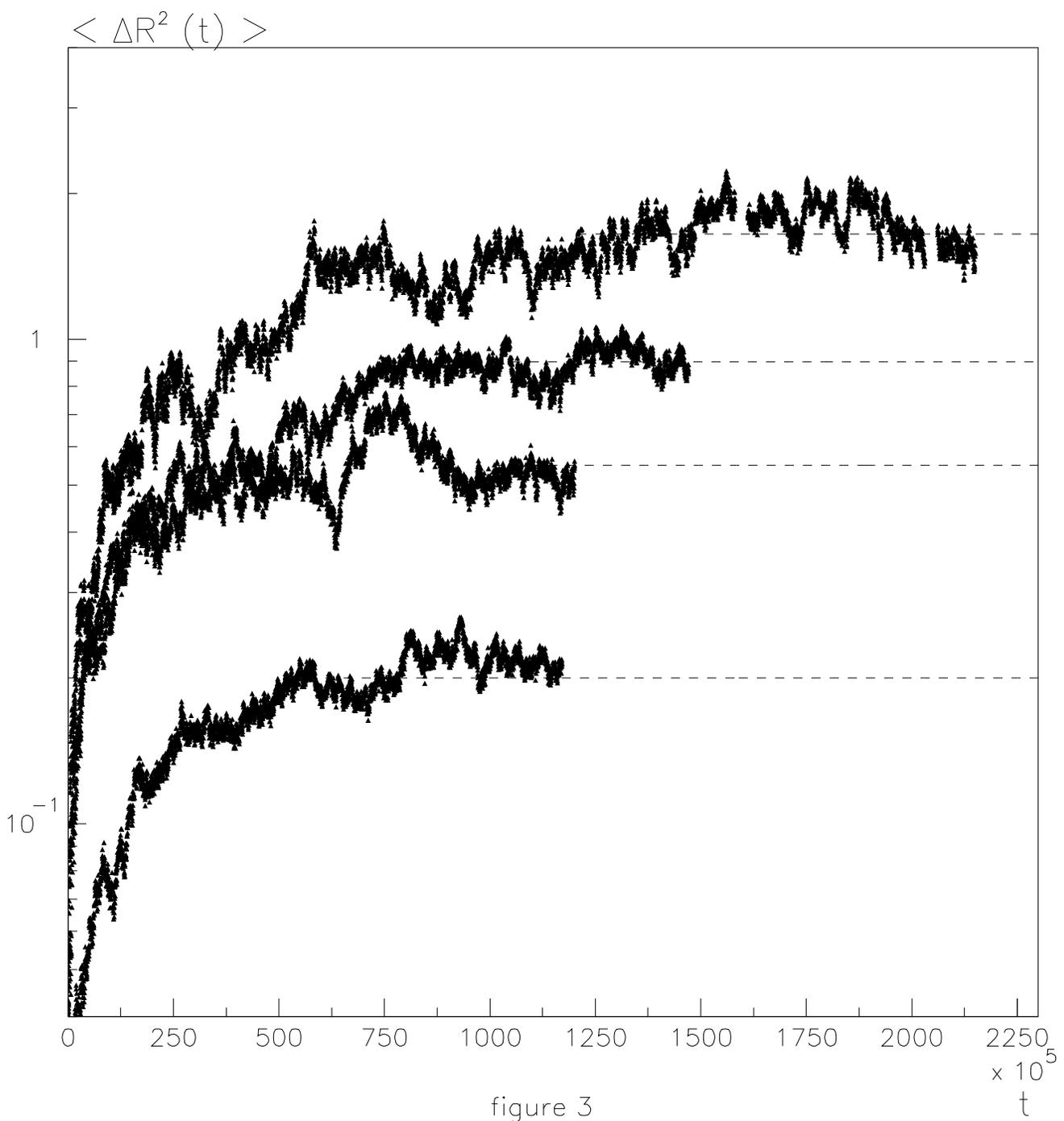
\begin{figure}
\caption{The mean square displacement of the center of mass of the percolating
cluster $\langle \Delta R^{2}(t) \rangle$ as function of time for different 
value of the monomer concentration $p$: from below $p=0.8,0.77,0.76,0,75$. The 
dashed lines correspond to the asymptotic plateau values $\Delta$ which grows 
approaching $p_{c}$. The data refer to a lattice size $L=32$. The unit time 
here is the MonteCarlo step per particle.} 
\label{fig3}
\end{figure}

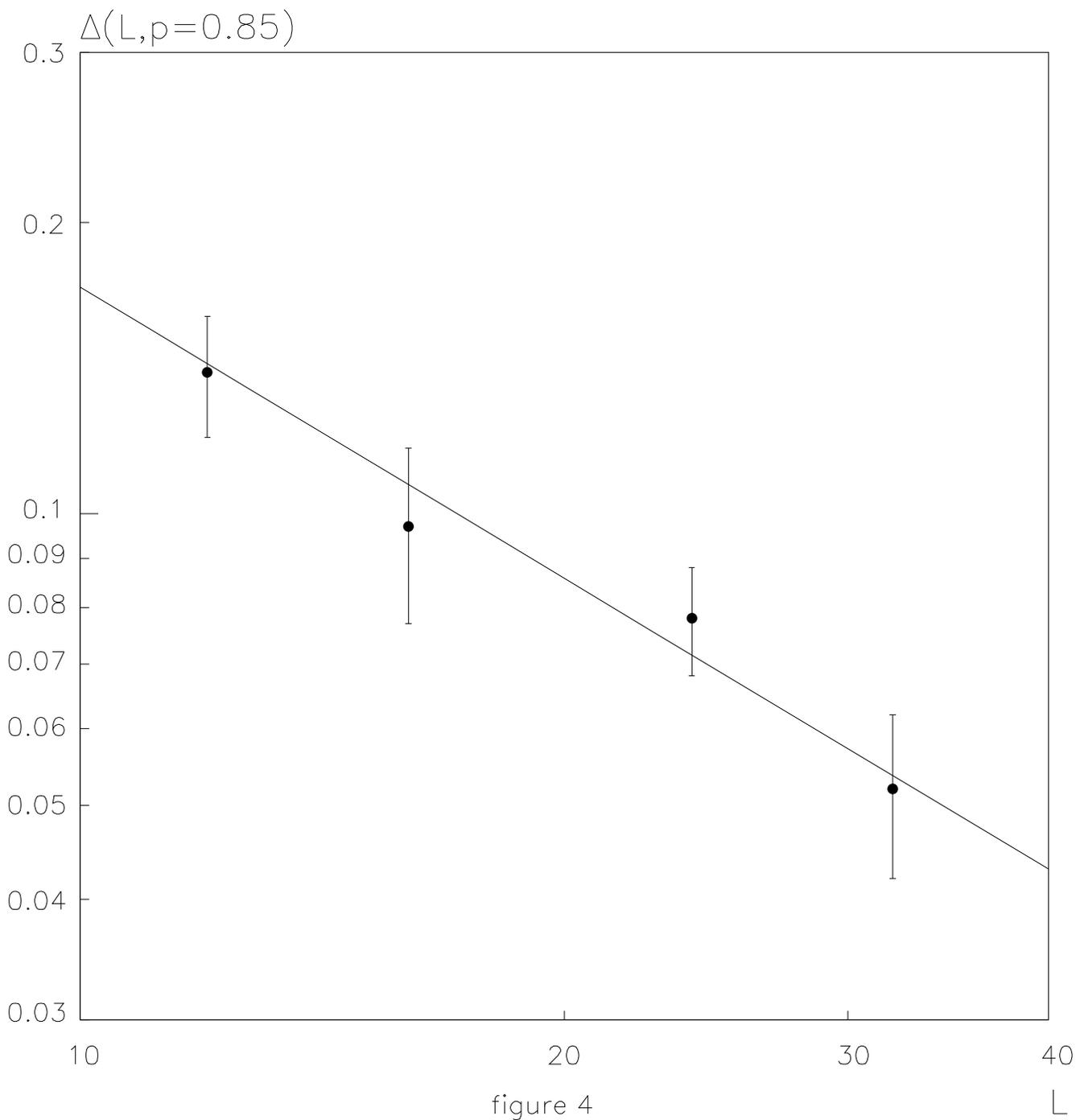
\begin{figure}
\caption{Log-log plot of $\Delta(L,p)$ for $p>> p_{c}$ (the data refer
to $p=0.85$) as function of $L$: the data show a scaling behaviour 
$\sim 1/L^{0.99 \pm 0.1}$. 
}
\label{fig4}
\end{figure} 

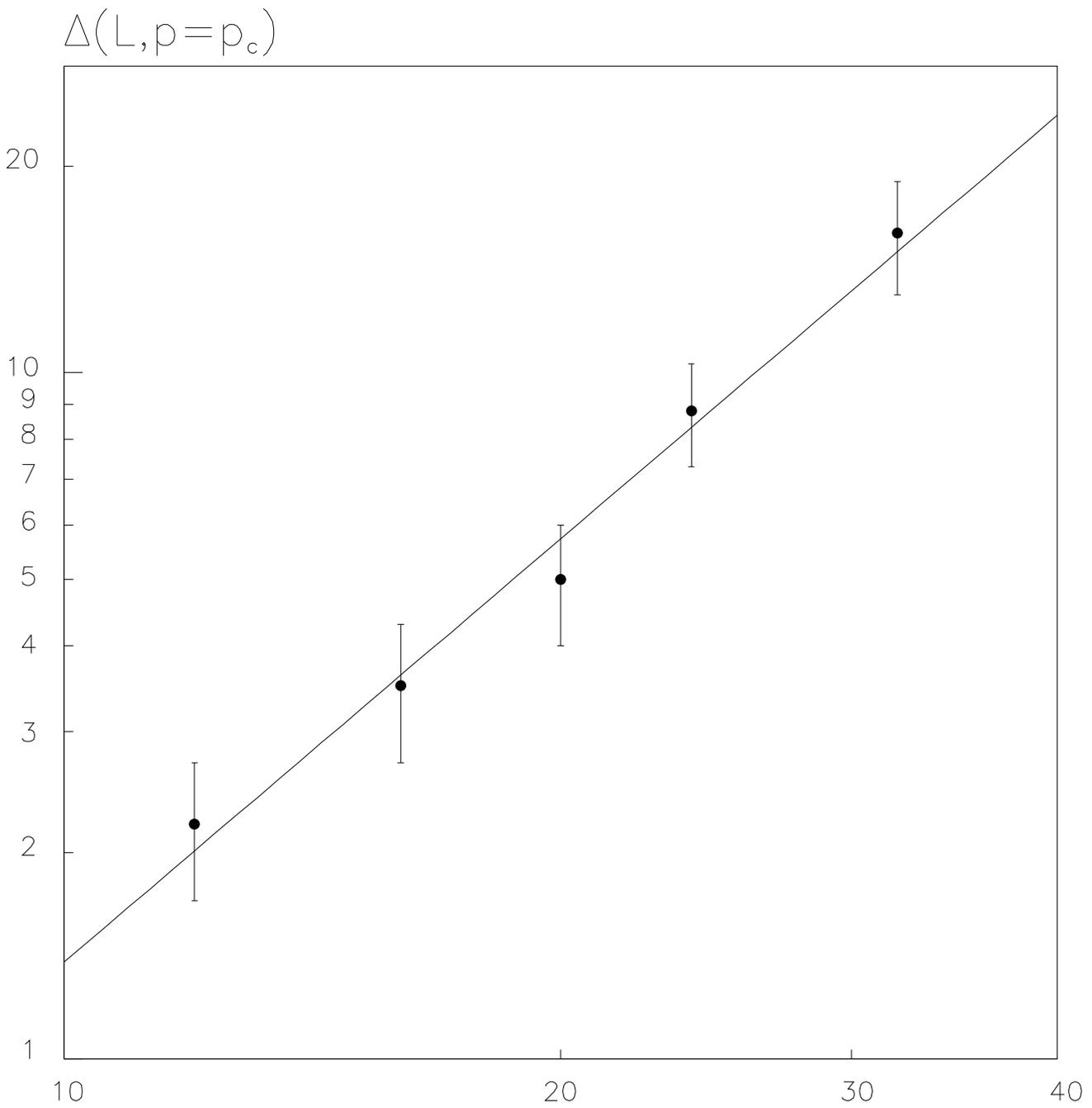
\begin{figure}
\caption{Log-log plot of the plateau values $\Delta(L,p=p_{c}=0.718)$ as 
function of $L$: the data are fitted by a power law giving the critical 
exponent $\tilde{z} \sim 2.0 \pm 0.1$. 
}
\label{fig5}
\end{figure}
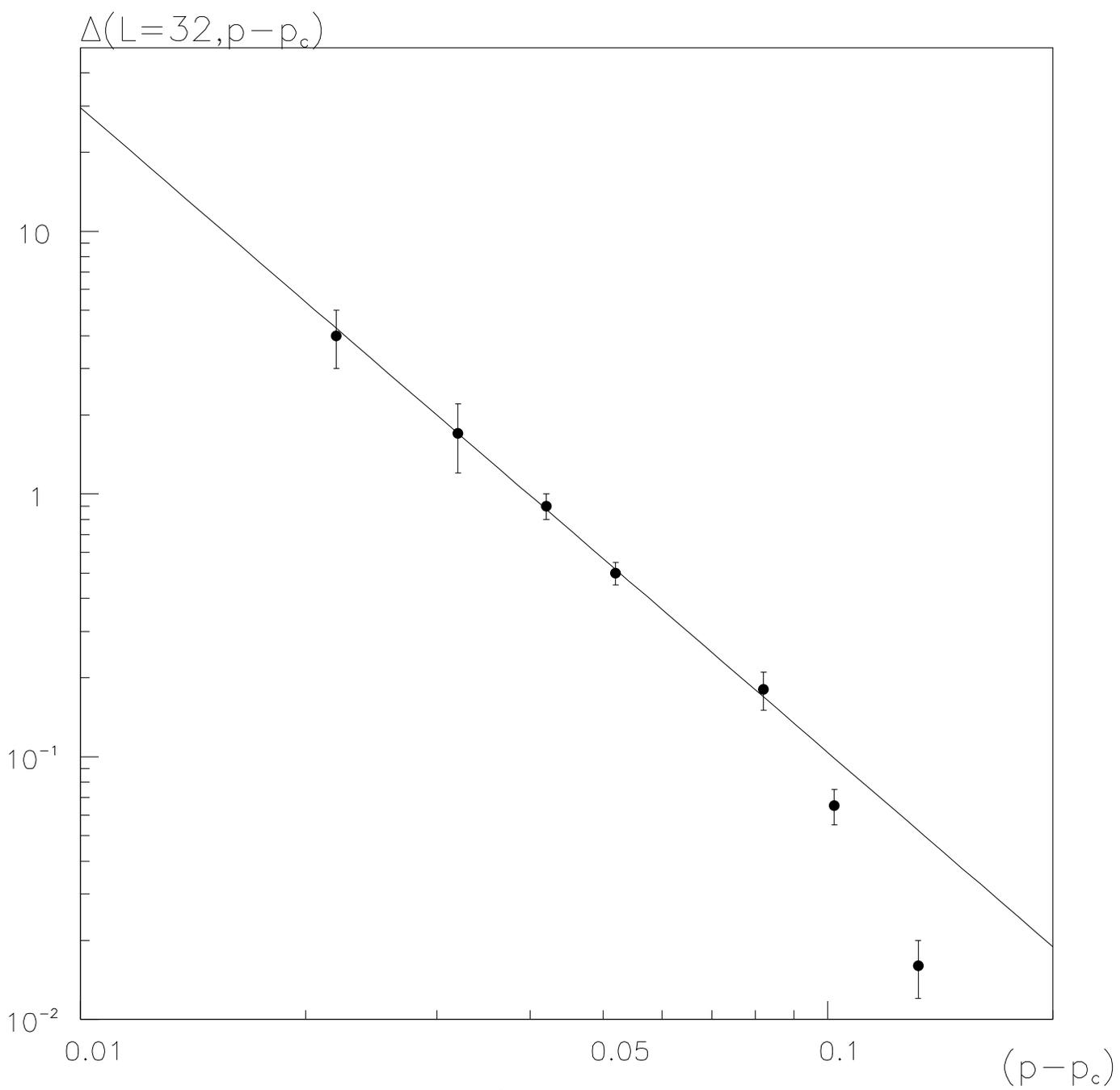
\begin{figure}
\caption{The plateau values $\Delta(L=32,(p-p_{c}))$ in a double logarithmic
plot as function of $(p-p_{c})$: the best fit of the data close to the 
percolation threshold gives the
critical exponent $f \sim 2.5 \pm 0.1$. 
}
\label{fig6}
\end{figure}


\begin{references}
\bibitem{note}     

Above the percolation threshold the percolating cluster can be considered 
as a network made of nodes connected 
by macrolinks of linear dimension $\xi$. Each macrolink can be considered as a 
spring of elastic constant $k$. For a system of size $L$ it can be regarded 
as $L/\xi$ elements in series, each element being made of $(L/\xi)^{d-1}$ 
springs in parallel. The macroscopic elastic constant is then 
$K = k(L/\xi)^{d-1} (L/\xi)^{-1}$.
Let $z$ be the critical exponent for the elastic constant $k$ of
the links of length $\xi$ decreasing to zero at the percolation threshold with
a power law behaviour $k\sim(p-p_{c})^{z}$ \cite{conda,kawe,staul}.
Then we obtain a scaling
behaviour for the macroscopic elastic constant $K \sim (L/\xi)^{d-2}
\xi^{-\tilde{z}}$, where $\tilde{z}=z/\nu$. This means that $K \sim
\xi^{-\tilde{z}-(d-2)} L^{d-2}$ and therefore again 
$K \sim L^{d-2} (p-p_{c})^{z+(d-2)\nu}$.
\bibitem{note2}

In both approaches we stress that the fluctuations for the percolating 
clusters have been calculated in presence of the finite clusters and the other
free monomers dynamically evolving according to the same fluctuation dynamics,
 namely we have studied the dynamic evolution of the percolating cluster 
embedded in the medium formed by the finite clusters, free monomers and 
solvent.
\bibitem{note3}

With the hard-wall boundary conditions, or free boundary conditions, the
fluctuation of the center-of-mass motion of the percolating cluster, 
whose size is $\sim L$, corresponds to the macroscopic elastic response of the
system. 
With these boundary conditions the simulation is actually performed at 
constant volume, i.e. there is a force acting on the percolating cluster which 
confines it inside the volume $L^{3}$. Following the picture presented at
the end of section III, let us consider the very simple case of a  
spring of elastic constant $K$ in a 
thermal bath at temperature $T$. In order to take into account the 
effect of the boundary conditions on the system we are studying, 
let us consider a constant force of module $f$ acting on it along the direction
$\hat{x}$ of its elastic deformation and introducing in the energy a linear
term in the deformation $x$. The mean energy of the
system is now 
$\langle U \rangle = \frac{1}{2} K \langle x^{2} \rangle - \langle f x\rangle$.
 If we assume that $\langle f x \rangle \sim f \langle x \rangle$ 
the mean fluctuation of the energy is again given by $ \langle \Delta U
\rangle = \frac{1}{2} K \langle (x -\langle x \rangle)^2 \rangle$, 
with $\langle x \rangle = f/K $. On the other hand
$\langle x \frac{ \partial U}{\partial x} \rangle = K_{B}T =  K\langle x^{2} \rangle - 
f \langle x \rangle$ and with the assumption 
$\langle f x \rangle \sim f \langle
x \rangle$ gives $K(\langle x^{2} \rangle - \langle x \rangle^{2}) = K_{B}T$. 
Then again $\langle \Delta x^{2} \rangle = 
(\langle x^{2} \rangle - \langle x \rangle^{2}) \sim 1/K$.
\end{references}
\end{document}